\documentclass[conference]{IEEEtran}
\usepackage[utf8]{inputenc}
\usepackage{graphicx}
\usepackage{fancyhdr}
\graphicspath{ {images/} }
\title{ECrits - Visualizing Support Ticket Escalation Risk}

\author{
    \IEEEauthorblockN{Lloyd Montgomery, Emma Reading, Daniela Damian}
    \IEEEauthorblockA{\textit{SEGAL Lab, University of Victoria}\\
                      Victoria, Canada\\
                      \{lloydrm, readinge, danielad\}@uvic.ca}}
                      
\begin{document}

\maketitle

\begin{abstract}
Managing support tickets in large, multi-product organizations is difficult. Failure to meet the expectations of customers can lead to the escalation of support tickets, which is costly for IBM in terms of customer relationships and resources spent addressing the escalation. Keeping the customer happy is an important task in requirements engineering, which often comes in the form of handling their problems brought forth in support tickets. Proper attention to customers, their issues, and the bottom-up requirements that surface through bug reports can be difficult when the support process involves spending a lot of time managing customers to prevent escalations. For any given support analyst, understanding the customer is achievable through time spent looking through past and present support tickets within their organization; however, this solution does not scale up to account for all support tickets across all product teams. ECrits is a tool developed to help mitigate information overload by selectively mining customer information from support ticket repositories, displaying that data to support analysts, and doing predictive modelling on that data to suggest which support tickets are likely to escalate.
\end{abstract}

\section{Introduction}
Support personnel manage support tickets as a means of listening to customer concerns; these concerns often relate to bugs in the product, or translate to product enhancements and requirements. However, a large portion of their time is spent managing potential and ongoing escalations of those support tickets, instead of handling the underlying bottom-up requirement being presented by the customer. Escalations are a process initiated by unsatisfied customers who are looking for a faster or more thorough solution to their problem, but instead, escalations introduce more process, people, and resources directed at handling the escalation itself instead of the underlying issue behind the support ticket.

Support analysts, tasked with managing support tickets and preventing escalations, have to rely on the information provided to them in support tickets; however, support tickets often only contain information immediately relevant to the issue at hand such as a description of the issue, issue side-effects, resources affected by the issue, etc. Companies such as IBM collect and archive their support tickets across all offered products, creating a wealth of information. Support personnel within companies already have access to support ticket records, yet parsing and summarizing any amount of this data is a time-consuming task. Even if available data was collected and summarized, what conclusions are to be drawn from this data, what question should be answered? For our industry collaborator, that question is ``how likely is this support ticket to escalate?" To answer that question, more than just the data relevant to this support ticket are needed, trends in the entire data set of support tickets may be relevant.

Given that a company collects and archives their support ticket data, scripts can be written to collect and summarize data relevant to the support ticket at hand, saving support personnel many hours of tedious work. Once collected, that data can be displayed to support personnel so they can draw their own conclusions about the state of the data and the implications it has for the support ticket. In addition to that, the collected data can be used to build a predictive model to provide support analysts with a quick-reference initial-triage summary of the risk of escalation for each support ticket.

\section{ECrits}
ECrits is a communication and issue-tracking tool that allows users to track support tickets, manage PMR escalations, and communicate with other team members regarding them. At IBM, support tickets are called Problem Management Records (PMRs). The tool was developed iteratively in collaboration with support analysts at our industrial partner IBM. Initial prototypes of the tool were used and tested for usability during daily support management meetings over a period of four weeks and features suggested by the analysts were implemented incrementally.

ECrits builds on our work on Machine Learning (ML) techniques for escalation prediction to support IBM support analysts in real time assessment of escalation risk \cite{montgomeryRE2017}. The ML technique used to be the model used in this tool is a Random Forest classifier fed a number of engineered features built from the data collected in their ecosystem. While our research paper describes the model and the engineered features, here we describe ECrits: a tool that delivers the results of the model in an actionable form.

ECrits has two main views, the Overview and the In-Depth view. The Overview allows support analysts to view all of the active PMRs in their organization, with some limited information being displayed about each PMR. The Overview also allows support analysts to ``follow" PMRs they wish to see at all times in the sidebar. The In-Depth view contains the information for one PMR, with a much more detailed accounting of all the available data for that particular PMR.

\section{Motivation}
\subsection{Initial Assessment of New PMRs}
Assessing new PMRs is the first step in the issue-management process, the first challenge of which is gathering and summarizing available data on the customer and issue. Often, all available data on the \textit{issue} is included in the PMR, however, all available data on the \textit{customer} is spread across company archives, most notably across other PMRs. The algorithms behind ECrits gather relevant customer information from all other PMRs in the system and calculates customer profile metrics. Those metrics are then fed into the ML model as features and subsequently displayed within ECrits as seen in Fig. \ref{fig:indepth}. Displaying these attributes, along with the other three categories, grants support personnel additional insights into the tickets at hand that they would otherwise have to find and calculate themselves.

\begin{figure}[t]
    \centering
    \frame{\includegraphics[width=\columnwidth]{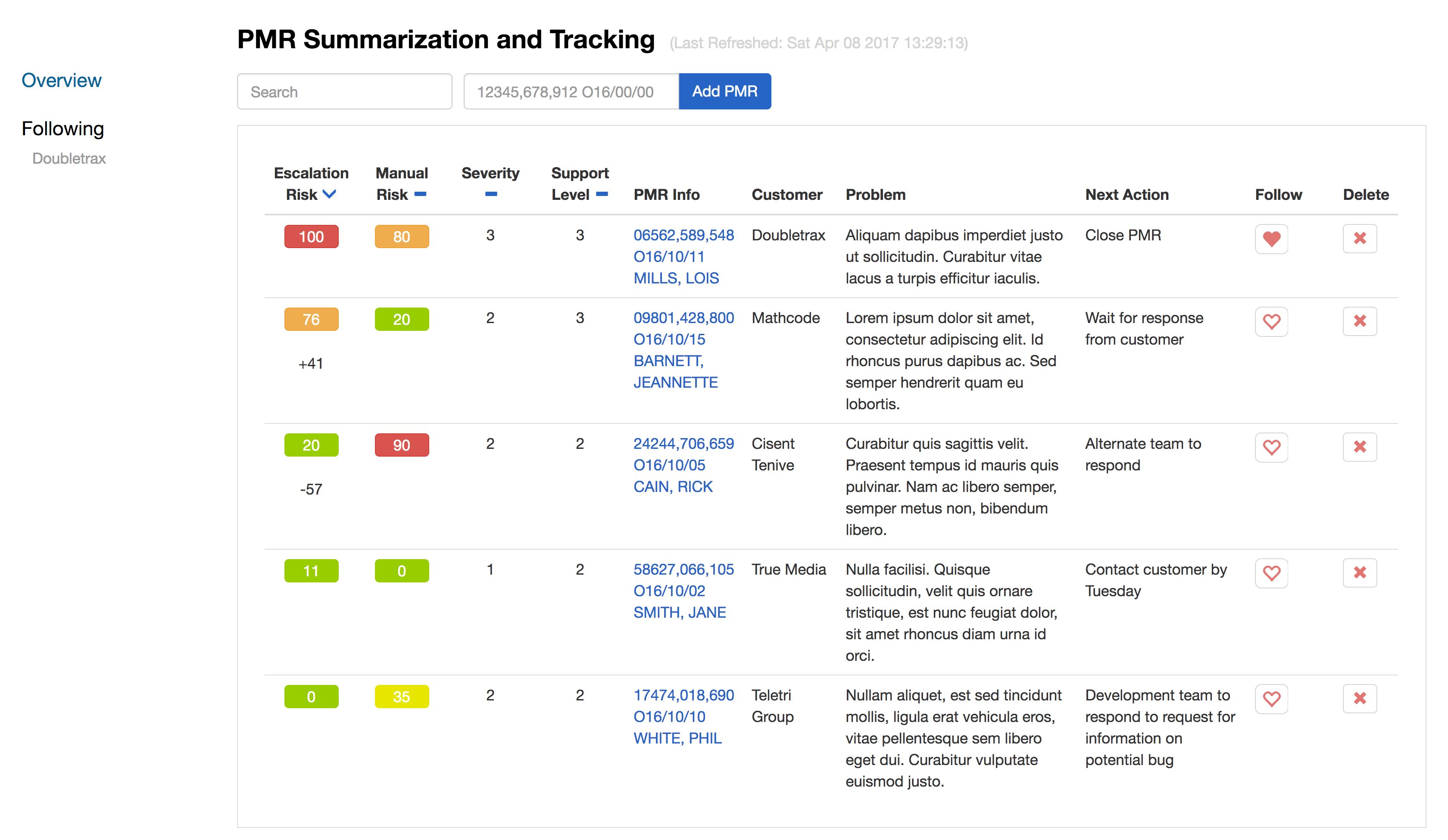}}
    \caption{ECrits Overview Page}
    \label{fig:overview}
\end{figure}

\subsection{Investigation of PMR Updates}
Support analysts may check on their PMRs at any time, and knowing what has updated since their last check is the most pertinent information. ECrits displays a Change in Escalation Risk if the Escalation Risk has changed since the last update. This is a visual notification that something within the PMR has changed, and, the impact it had on the likelihood of escalation. ECrits displays live data from IBM's support-ticket ecosystem, and the ML results are updated at certain intervals throughout the day. At any time, IBM support analysts can check the tool and see what these values are. Together, the ML results and the underlying features that are displayed provide support analysts with a quick method of checking updates to their PMRs.

\subsection{Collaboration Through Communication}
\begin{figure}[t]
    \centering
    \frame{\includegraphics[width=\columnwidth]{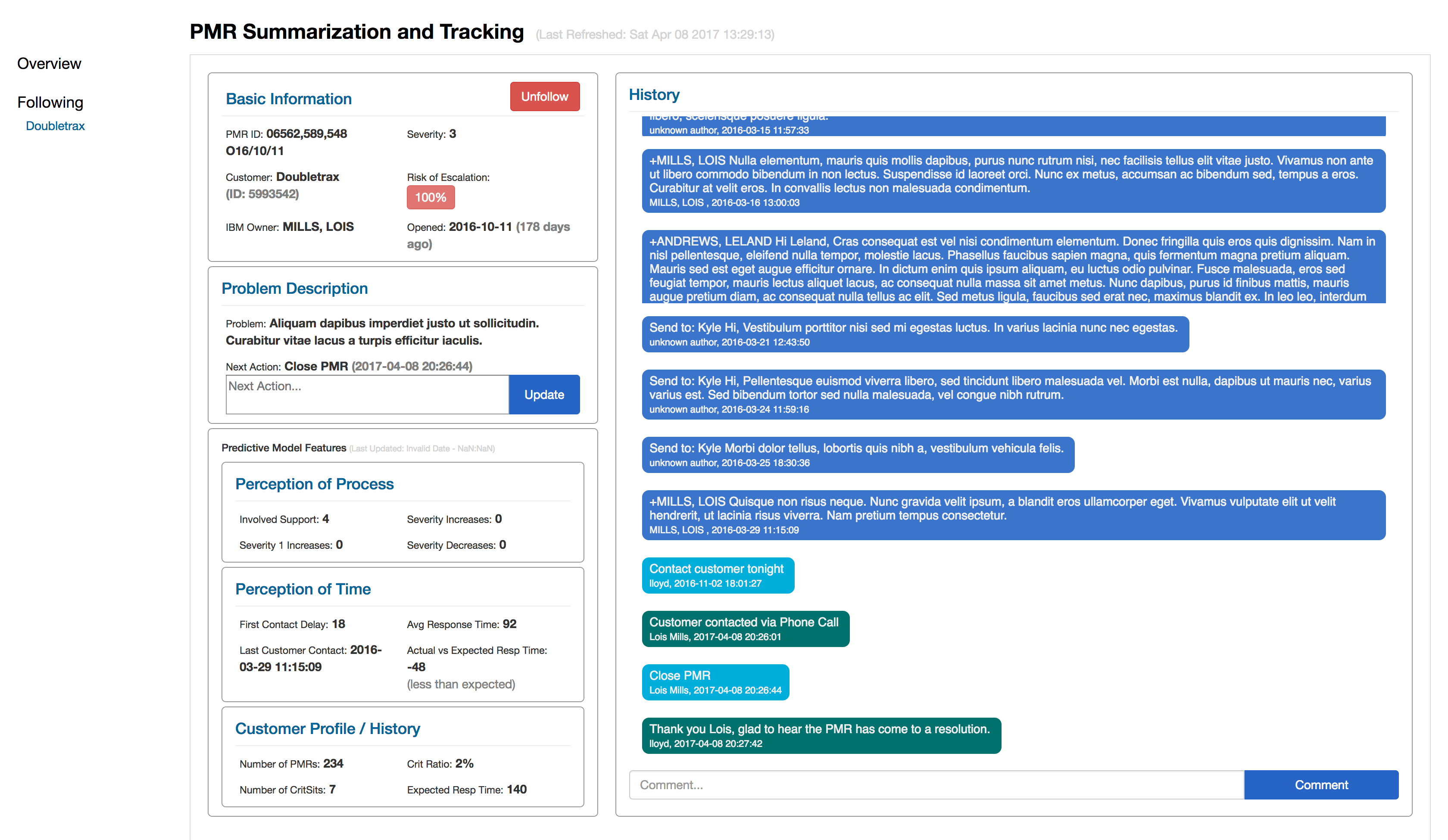}}
    \caption{ECrits In-Depth View for A PMR}
    \label{fig:indepth}
\end{figure}

Previous to ECrits, team members communicated via email or in person and their conversations were not saved directly to the PMR they were communicating about. As multiple support analysts may work on a single PMR throughout its life cycle, keeping track of all changes and updates to a PMR is very important. ECrits maintains a comment system on individual PMRs that weaves comments in with the updates, as seen in Fig.  \ref{fig:indepth} as green text blobs. In addition to the comment ability, ECrits has a ``next action" collaboration feature where the next action reflects the next actionable task on the PMR, displayed as light blue text blobs. This textual field is filled in and submitted similar to a comment in that it temporally weaves itself into the updates and comments of individual PMRs, but in addition to that, the next action displays in the Overview page (Fig.  \ref{fig:overview}) to allow collaborators to easily check and follow-up on the completion status of a current next action.

\section{Implementation}
\subsection{Technologies}
Ecrits is built on a modified MEAN stack using MySQL, ExpressJS, AngularJS, and NodeJS. The MEAN stack allows the entirety of the web application to be built using JavaScript. NodeJS is a server-sided JavaScript framework and ExpressJS is a NodeJS framework that makes it easier to manage endpoints within NodeJS. AngularJS is a client-side front-end JavaScript web framework and was used for building the dynamic UI for Ecrits. Finally, MySQL is a relational database that was used for storing user information and communication data.


\section{Conclusions}
Machine Learning applied to support datasets and visualized for support personnel is a field of study that needs more examples of practical applications. ECrits provides such an example through a collaboration with IBM using real data and deployed within their production environment. ECrits also provides a novel approach to managing and utilizing customer data in the support process: the data used to create the escalation risk is directly displayed to support analysts so they can come to their own conclusions about the risk of escalation, and weigh that against the predicted risk.

The future improvements of this tool include providing a more detailed account -- per PMR -- of the data used in the model, as well as improving the underlying ML model to produce more precision predictions. We are also looking to perform a longer study at IBM to record and compare metrics to test the hypothesis that a tool such as this provides a notable increase to actionable situations through awareness of the data. 

\bibliographystyle{IEEEtran}
\bibliography{main}

\begin{thebibliography}{1}
\providecommand{\url}[1]{#1}
\csname url@samestyle\endcsname
\providecommand{\newblock}{\relax}
\providecommand{\bibinfo}[2]{#2}
\providecommand{\BIBentrySTDinterwordspacing}{\spaceskip=0pt\relax}
\providecommand{\BIBentryALTinterwordstretchfactor}{4}
\providecommand{\BIBentryALTinterwordspacing}{\spaceskip=\fontdimen2\font plus
\BIBentryALTinterwordstretchfactor\fontdimen3\font minus
  \fontdimen4\font\relax}
\providecommand{\BIBforeignlanguage}[2]{{%
\expandafter\ifx\csname l@#1\endcsname\relax
\typeout{** WARNING: IEEEtran.bst: No hyphenation pattern has been}%
\typeout{** loaded for the language `#1'. Using the pattern for}%
\typeout{** the default language instead.}%
\else
\language=\csname l@#1\endcsname
\fi
#2}}
\providecommand{\BIBdecl}{\relax}
\BIBdecl

\bibitem{montgomeryRE2017}
L.~Montgomery and D.~Damian, ``What do support analysts know about their
  customers? on the study and prediction of support ticket escalations in large
  software organizations,'' in \emph{Proc. of IEEE International Conference on
  Requirements Engineering}, August 2017, to appear.

\end{thebibliography}

\section*{Additional Attachment Content}

\section*{Overview}

For the tool demo presentation we will have a working copy of the web-app tool running locally on a laptop. This live demo will feature mock data (since real data has privacy concerns) that mimics what the tool would look like if it was running in a real support environment. We also have a poster available we can bring that highlights important aspects of the tool and how it works, thereby adding to the overall live demo setup. This poster \textit{is not} part of the submission, but we can bring it to display behind our demo if it is a welcomed addition to our display. Let us know if that is recommended.

This demo, featuring mock data, can be interacted with by participants, including adding their own comments, next actions, and investigating the data behind the predictive values. We encourage participants to explore the answers to questions such as ``why is the escalation risk 100\%?" and "how come the estimated escalation risk is so far off from the predicted risk?". This investigative process is what we hope the tool encourages.

\section*{Demo Walk-through}

Participants will be encouraged to use the tool to investigate the support tickets, escalation risks, and machine learning attributes, as well as be guided through the purpose of each screen available to them as mock-analysts. The presenter will explain how the machine learning attributes are connected to the predictions and escalation risks, and then ask participants to comment on the connection they see in the tool.

The tool, as explained in the paper, only consists of two views: the Overview and the In-Depth view (there is different data per PMR in the In-Depth views). The simplicity of the tool means that participants will not spend long learning about the tool or its use-cases, but rather just viewing and reasoning about the data. 

The purpose of the tool is to deliver data to support analysts, providing them with additional insights into the support tickets they are working with. Particularly with regards to predicted values and the data that was used to build them. This demo will provide participants with the same experience, allowing them to see support tickets, data, predictions, and connections between predicted values and machine learning features.

\section*{Information}

We can be reached at lloydrm@uvic.ca and danielad@uvic.ca for questions and comments regarding the tool.

\begin{figure*}[t]
    \centering
    \frame{\includegraphics[width=\textwidth]{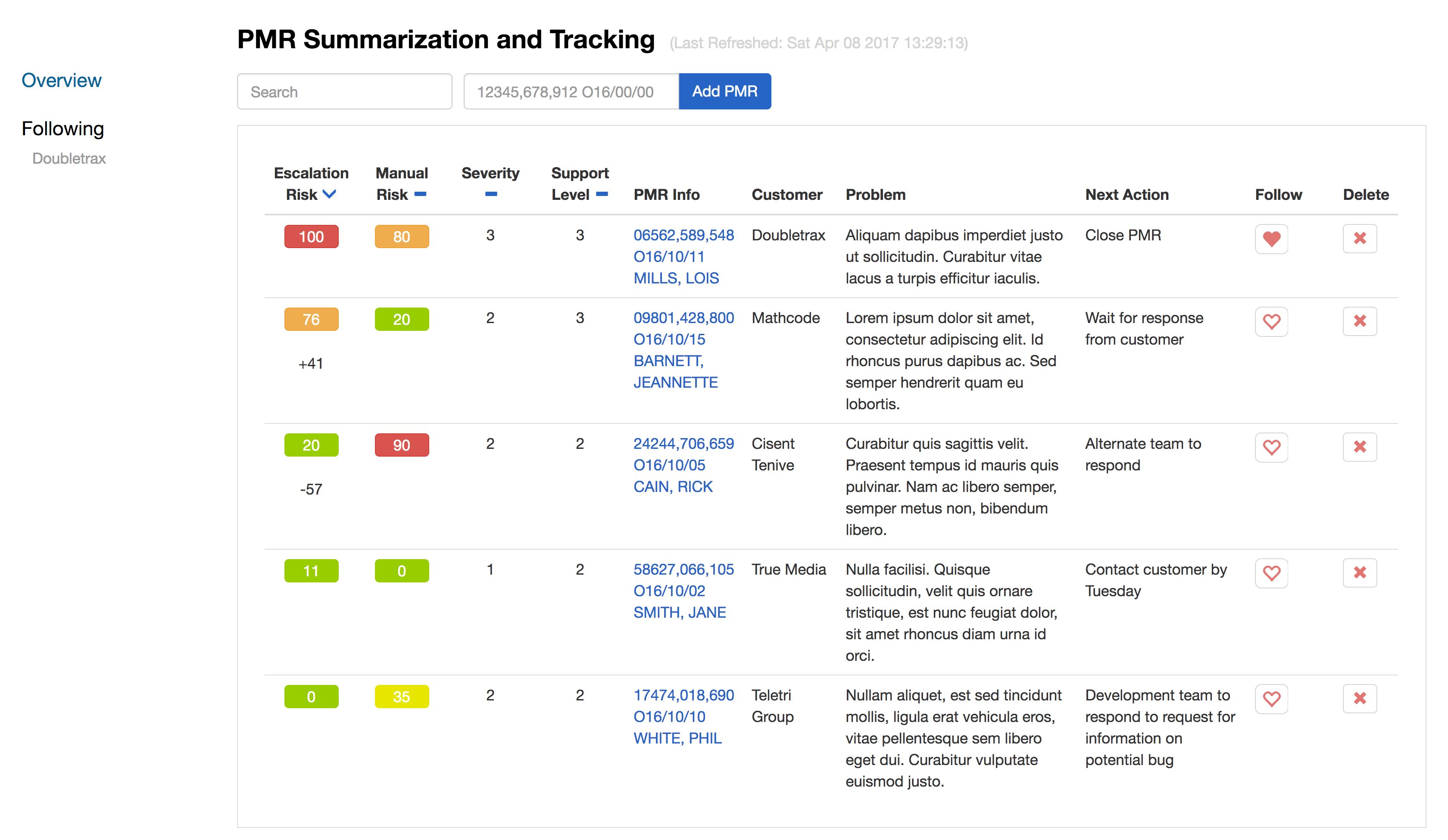}}
    \caption{ECrits Overview Page -- Bigger}
    \label{fig:overview_big}
\end{figure*}
\begin{figure*}[t]
    \centering
    \frame{\includegraphics[width=\textwidth]{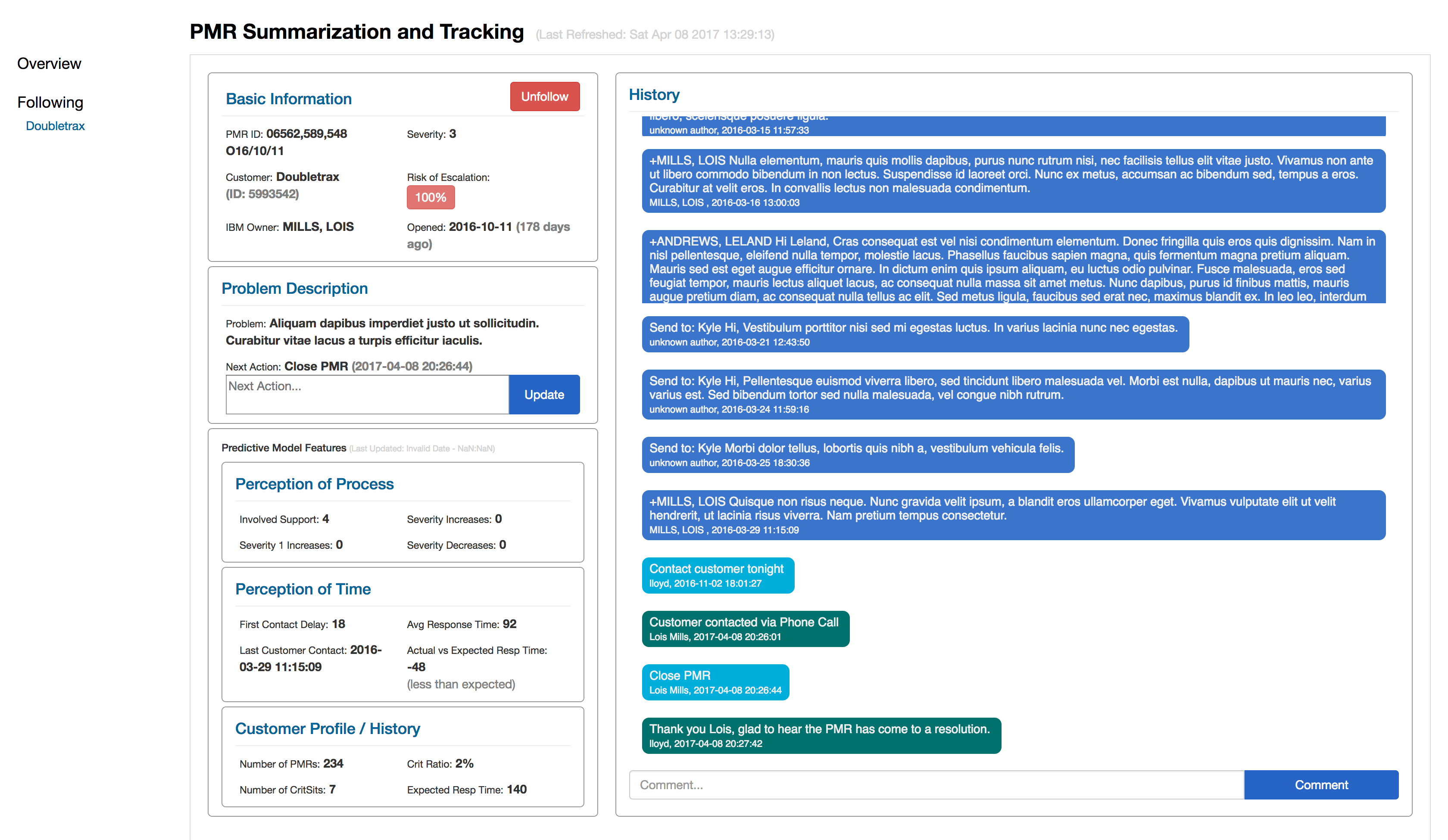}}
    \caption{ECrits In-Depth View for A PMR -- Bigger}
    \label{fig:indepth_big}
\end{figure*}

\end{document}